\documentclass[twocolumn, showpacs, amsmath, amssymb, aps, prl]{revtex4-1}
\usepackage{graphicx}
\usepackage{dcolumn}
\usepackage{bm}

\begin{document}

\title{Quantum motor and future}

\author{Evgeny\,G.\,Fateev}

 \email{e.g.fateev@gmail.com}
\affiliation{%
Institute of mechanics, Ural Branch of the RAS, Izhevsk 426067, Russia
}%
\date{\today}

\begin{abstract}
In a popular language, the possibilities of the Casimir expulsion effect are 
presented, which can be the basis of quantum motors. Such motors can be in 
the form of a special multilayer thin film with periodic and complex 
nanosized structures. Quantum motors of the type of the Casimir platforms 
can be the base of transportation, energy and many other systems in the 
future. 
\end{abstract}

                           
\maketitle

Well, the epoch of wheels and roads, petrol and rocket engines and other 
things that seem to be promising like thermonuclear reactor, is coming to 
its end{\ldots} Recently, the possibility of the construction of propulsive 
systems based on the quantum effect of Casimir expulsion has been shown 
theoretically. More detailed information about the effect can be found in 
papers published in the electronic archives 
\cite{Fateev:2012a, Fateev:2012b, Fateev:2013}. And it 
is not something supernatural or unreal; it is something that can be 
realized in the nearest future. Let us briefly note that if periodic 
structures based on Casimir expulsion are made from almost ideally 
conducting metals (such as gold and platinum) and for the ideal conditions, 
one layer of such structures with the area 6 -- 8 square meters could lift 
Cheops pyramid (6.25 million tons)! And not just lift it but it could set it 
in motion and push it to the ends of the Universe at the light speed if we 
wish. It is clear that much more serious and universal quantum effect than 
levitation (i.e. keeping one body over the other) or any abstract 
antigravitation (gravitation repulsion so much talked about) is at issue.

Specifying the notion of expulsion we should note that it is associated
with a constant moving force in contrast to repulsion (levitation effects)
and it does not require the presence of a partner body. It needs only the
presence of a material medium in which such effect occurs in a configuration.
Expulsive forces also include the buoyancy force which is the result of the
action of a gravitational field in a medium. This force acts upon closed
cavities in a material medium causing their movement in the direction
opposite to that of the gravitational field vector to the boundary where
gravitational and buoyancy forces get balanced. Unlike buoyancy, the
Casimir expulsion can act in a space directionally and constantly upon
open cavities because the boundaries of a vacuum medium are specified by
the cavity configuration itself. A particular case of such configurations
is Casimir parallel mirrors \cite{Casimir:1948, Casimir:1949, Milton:2001,
Bordag:2009, Bordag:2009a} which do not possess an 
effective expulsion force \cite{Fateev:2012a, Fateev:2012b, Fateev:2013}.
The uniformity of the action of such forces upon the configuration can
surely be changed at gradual or abrupt change of the parameters of the
material medium. 

In reality, most likely due to the existence of various unexpected 
obstructive physical factors, multilayered structures (10 -- 100 layers) 
will be needed. However, if we take into consideration that one layer of 
quantum structures will not be thicker than 10 angstroms, a multilayered pie 
of structures designed for lifting Cheops pyramid and its flight in the 
Universe will not exceed 10 -- 100 nanometers. It is 10 000 times smaller 
than the diameter of a human hair! Not too much for the flight of the 
pyramid. And for the flight of a man, a layer of the structures with the 
area less than 10 square millimeters will be required with regard for 
possible corrections for roughness of internal elements of the structures 
and other possible problems! For providing 7 billion inhabitants of the 
Earth with the possibility of flying, in general less than 10 -- 100 
kilograms of gold will be necessary!!! By the way, aluminum can be used as 
well{\ldots} Without going into the specific constructive peculiarities of 
making such structures, let us envisage spheres of their use and grand 
perspectives in the future that will be opened to the humanity:

\begin{enumerate}
\item \hyperlink{1}{Transport}
\item \hyperlink{2}{Energy}
\item \hyperlink{3}{Medicine}
\item \hyperlink{4}{Astronautics}
\item \hyperlink{5}{Robot technologies}
\item \hyperlink{6}{Home appliances}
\item \hyperlink{7}{Construction and Architecture }
\item \hyperlink{8}{Food replication}
\item \hyperlink{9}{World ecology}
\item \hyperlink{10}{Life safety systems}
\item \hyperlink{11}{Life philosophy}
\end{enumerate}

\hypertarget{1}{\textbf{1. Transport}}

Panels with quantum Casimir effect of \textbf{expulsion} can be the basis of 
propulsive forces of all transportation platforms. Panels can be arbitrarily 
powerful and compact. In this sphere, there is a great field for creativity 
for inventors. Transport facilities can be of any size and lifting capacity. 
Nowadays, car frames, wheels and engines make up about 70 -- 90{\%} of the 
vehicle weight. They will become unnecessary. It is possible to make 
individual transportation capsules capable of ``hovering'' over the ground 
at any height and in any direction needed. The need for roads with hard 
asphalt surface or any other surface between cities and settlements will 
disappear. It means that roads can be forgotten forever. Released lands will 
be reclamated and turned into lawns and fields. In cities, the noise of 
transport will disappear. The air will become clean. Production of oil, gas 
and other mineral products used for the creation and operation of modern 
means of transport will become unnecessary. The negative influence of the 
human activity on the environment will be minimized.

\hypertarget{2}{\textbf{2. Energy}}

The production of electricity required for the operation of electric 
apparatus and electronics will be realized on the basis of the Casimir 
expulsion effect. The electricity production will take place directly where 
electric and electronic devices are set. Devices will be compact enough and 
powerful. The necessity for the expensive means of electricity supply such 
as high- and low-tension power lines will disappear.

Humanity will have no need for powerful electricity generating stations, 
such as atomic power stations, thermal power stations, water power stations, 
tidal power stations, solar power plants, wind power stations, etc. The 
development of hydrogen power engineering and everything associated with it 
will become pointless. It will become unnecessary to develop and construct 
thermonuclear power plants. Consequently, it will be unnecessary to make 
electric generators and other equipment for these power plants. The need for 
providing such object with expensive security systems will disappear as 
well.

The closure of all types of power plants and stations, especially atomic 
ones, can remarkably improve the planet ecology. It will lead to the 
complete cessation of the destructive action of the waste of the 
commercial-scale power production on the nature. People working in the 
sphere of energy production will be able to use their knowledge and 
experience in other industries, for example, such as nano-production 
corporations or the expansion of humanity into the space.

\hypertarget{3}{\textbf{3. Medicine}} 

There are innumerable possibilities for the application of the Casimir 
expulsion effect in medicine. Since the devices for the energy 
transportation and generation on the basis of the effect can be arbitrarily 
small, there will be the possibility of the creation, for example, of a 
supercompact artificial \textbf{heart} continuously operating without 
batteries and wires, robotic nano-devices for \textbf{cleaning} and 
restoration of blood vessels and organs, for balancing blood \textbf{sugar} 
and ``hunting'' for \textbf{malignant} cells. The effect can be used for 
creating donor organs.

Special nanorobots will make large-scale surgical operations, for example, 
on the elimination of acquired or inherited defects in organs. The Casimir 
expulsion will be used in stomatology. For example, dental prostheses 
levitating near gums will be created. For positioning dental prostheses, 
magnetic marks implanted into a gum will be used. The firmness of the 
adjacency of dental prostheses to gums will be regulated automatically 
depending on the task performed by them at the particular moment. Traumatic 
technologies of fixing artificial teeth in jawbones will become things of 
the past.

\hypertarget{4}{\textbf{4. Astronautics}}

There is the possibility to create multilayered structures on the basis of 
the Casimir expulsion having \textbf{arbitrarily large total thrust}. The 
structures will have small weight, many times smaller than the lifted weight 
of the sheath and content of space vehicles. There is no need to mention 
that the systems will be repeatedly used; they are reusable on default. It 
is possible \textbf{to create spacecrafts weighing millions of tons} with 
arbitrarily powerful protection against asteroids and other dangers.

Highest possible speed of motors on the basis of the expulsion effect can be 
the velocity of light. Such motors will not have disadvantages which are 
inherent in motors used nowadays. For example, the lift of the apparatus 
from any unequipped place will be possible. Even motors on the basis of 
photon radiations seem archaic, senseless and unreal. The expansion of the 
humanity into the boundless space is possible only with the use of motors 
based on the Casimir expulsion. Naturally, visiting various cosmic objects 
and settling on neighboring planets will be possible in the nearest future 
(much sooner than it is being planned by modern cosmic agencies).

Let us consider the hypothetical possibility of visiting the Earth by 
aliens. In principle, it could be possible only with the use of motors based 
on the Casimir expulsion. Since the effect provides infinite energy and 
perfect maneuvering capabilities, it is doubtful that aliens' spacecrafts 
could have any wrecks on the Earth. And if anything had taken place due to 
any strange concurrence of circumstances, \textbf{some} of such apparatus or 
their parts would have perpetually ``drifted'' or hung like mountains over 
the Earth. We do not observe such phenomena. Consequently, it can be stated 
with 100{\%} probability that the Earth has not been visited by aliens 
during the entire history of its existence and they are not visiting the 
Earth nowadays.

\hypertarget{5}{\textbf{5. Robot technologies}}

In the future, robots will be different from those described in scientific 
fictions and demonstrated in movies. Let us say, science-fiction writers are 
not quite strong visionaries! The use of Casimir platforms most likely 
deprives robots of anthropomorphous features such as ``hands'', ``legs'' and 
even ``body''. Moreover, robots will not need to recharge. It means that for 
performing a certain mechanical function, a robot will have only 
platforms-\textbf{grippers} which will not be necessarily bounded 
mechanically with one another. As for the other parts of the ``body'', such 
robot will not have them in principal.

\hypertarget{6}{\textbf{6. Home appliances}}

Any home appliances and furniture will undergo certain changes. Each 
domestic device such as a refrigerator, vacuum-cleaner or computer will be 
equipped by an individual compact energy source on the basis of the Casimir 
effect of expulsion. Therefore, the devices will become wireless and operate 
silently. New universal appliances will appear such as regenerators of life 
support. They are devices which will regenerate air and water. Actually, it 
will be possible to live in any place of the space using such devices. That 
is, they are modules based on endless energy which will provide the humanity 
expansion in the Universe.

Furniture can be placed on platforms and pieces of furniture will become 
much lighter and more mobile. In fact they can be compact enough to be 
carried during walking tours.

Clothes and footwear will be modernized. Inserts with thin and flexible 
layers of Casimir platforms can be sewn in clothes. As the result, a human 
being or a domestic animal can gain transport possibilities of flying 
quality. Thus, putting on special socks, a person will be able to walk on 
the water surface. And putting on trousers and mittens one can learn to fly 
artfully in the atmosphere. In this case, everything will be provided that 
the flyer would never fall on the ground. A human being will gain fantastic 
possibilities for tourism and travels without additional flying vehicles and 
will be able to travel individually, with his or her family or with a group 
of people. Having special spacesuits it will be possible to fly individually 
or with a group to the space, the Moon and neighboring planets.

\hypertarget{7}{\textbf{7. Construction and Architecture}}

Since platforms based on the effect of vacuum expulsion can make buildings 
fly, heavy foundations and massive supporting structures will become 
unnecessary. The walls of buildings will be thin even in the areas with cold 
climate, because compact and inexpensive energy sources will appear, which 
will not burn oxygen. Each storey of a building will be easily dragged to 
any other group of buildings in a town or settlement. In fact, even lift 
cranes will become unnecessary because building storeys after they have been 
built will be able to occupy their positions at any height and any place. 
Reinforced concrete constructions will become unnecessary and reinforced 
concrete construction- and brick-making plants will be closed. Thus, 
buildings can be higher than clouds if there is such a need. Naturally, all 
buildings will be quakeproof. Wind and other loads will be damped 
automatically by side platforms. As it has been already mentioned, transport 
constrictions such as bridges will lose their practical meaning.

In connection with radical changes in the concept of construction, 
\textbf{architecture} will also change. Buildings will not be permanently 
located in a district. They will be capable of turning after the Sun and 
receiving sufficient amount of light rays. It will be possible not to take 
into account any wind loads. Water supply and sewage for each flat or 
building in general can be built into the cycle with the devices for 
regeneration of these resources. Each device will have wireless individual 
electricity supply. Such possibilities will urge the imagination of 
architects to original creative solutions, because strength and other limits 
will not influence them in their search of forms.

\hypertarget{8}{\textbf{8. Food replication}} 

Only the appearance of nanorobots moving and operating on the basis of 
vacuum forces of expulsion will make possible to create devices for the 
integration of food at the molecular level according to the program -- 
fruits, vegetables, eggs, meat, etc., and dishes based on them. They will be 
the so-called replicators - the embodiment of magic tablecloths. Millions of 
little nanorobots will be occupied with constructing and assembling 
molecules strictly according to the program for creating a certain product. 
The replica of such a product will not differ from its natural mate in 
anything. Since the energy of such vacuum platforms is endless, it will be 
possible to produce an infinitely large amount and variety of products.

\hypertarget{9}{\textbf{9. World ecology}}

The discovery of the effect of vacuum expulsion and its use will lead to a 
radically new jump in the development of the humanity. This development will 
be based on a much less use of the resources of the Earth. Industrial 
discharge of carbon dioxide into the Earth atmosphere will practically stop. 
The contamination of the Earth with nuclear waste will cease. Large 
territories will become free from highways and railways. Oil spill in seas 
and oceans will stop. In cities, noise pollution will decrease many times. 
Due to the cessation of the operation of hydropower plants and their 
dismounting, water bodies and rivers on the continents will return into 
their virgin state. All the above will improve the ecology of the world.

\hypertarget{10}{\textbf{10. Life safety systems}}

The use of the quantum effect of vacuum expulsion will make life of people 
safer. It will happen due to the improvement of the Earth ecology, the 
change of the entire infrastructure of life, the cancelling out of the 
content of military conflicts, because of the seismic and other safety of 
dwelling on the platforms of expulsion, the possibility of the 
deconcentration of dwelling houses over the planet territories.

Possible global cataclysms can be easily prevented. For example, 
trajectories of asteroids can be easily changed. Asteroids will be covered 
with platforms with the effect of vacuum expulsion and will be dragged to 
safe orbits. Moreover, all massive asteroids in the Solar System will be 
``caught'' by special ``trawls'' and removed beyond the System with the help 
of the vacuum platforms. Systems on the basis of our platforms will ``hang'' 
over each volcano and utilize ash and gases during volcanic eruption. In the 
oceans there will be platforms warming up or if necessary directing global 
water streams in the desired direction to create salubrious climate on the 
planet.

\hypertarget{11}{\textbf{11. Life philosophy}}

It looks like the humanity has stayed too long at the start. Countries and 
people on the Earth resemble beetles in a matchbox. In the matchbox, the 
pressure grows because there is no exit. Dissipative social processes 
develop into the struggle for a ``place in the Sun''. At best, the psychic 
energy of people is utilized and sublimated in sport and tourism; however, 
in the routine case, it is directed to leading a dissipative life, drug 
addiction and alcoholism. The possibilities of earthmen to show their worth 
in the space exploration are strongly limited. The creation of the apparatus 
for the expansion in the Universe will blow up the ``life in the matchbox'' 
and, finally, give a chance for each person to show his worth. A ``place for 
a deed'' will appear. Now people will be able to switch from the struggle 
for a place in the Sun over to the exploration of systems of stars. 
\begin{acknowledgments}
The author is grateful to T. Bakitskaya for hers helpful
participation in discussions.
\end{acknowledgments}


\begin{thebibliography}{8}
\bibitem{Fateev:2012a} E.G.Fateev, Casimir force of expulsion, \href{http://arxiv.org/abs/1208.0303}{arXiv:1208.0303v1} (2012).
\bibitem{Fateev:2012b} E.G.Fateev, Casimir expulsion of periodic configurations,
\href{http://arxiv.org/abs/1208.1256}{arXiv:1208.1256v1} (2012).
\bibitem{Fateev:2013} E.G.Fateev, Casimir expulsion of shifted configurations,
\href{http://arxiv.org/abs/1301.1110}{arXiv:1301.1110v1} (2013).
\bibitem{Casimir:1948} H. B. G. Casimir, Kon. Ned. Akad. Wetensch. Proc. \textbf{51}, 793 (1948).
\bibitem{Casimir:1949}H. B. G. Casimir, D. Polder, Phys. Rev. \textbf{73}, 360 (1948). 
\bibitem{Milton:2001} K. A. Milton, \textit{The Casimir effect: Physical manifestations of zero-point energy}, World Scientic, Singapore, 2001.
\bibitem{Bordag:2009} G. L. Klimchitskaya, U. Mohideen, and V. M. Mostepanenko, { Rev. Mod. Phys.} \textbf{81}, 1827 (2009).
\bibitem{Bordag:2009a} M. Bordag, G. L. Klimchitskaya, U. Mohideen, and V. M. Mostepanenko, \textit{Advances in the Casimir
 effect}, Oxford University Press, Oxford, 2009.
\end{thebibliography}
\end{document}